# ARTICLE

# 3D Printing in Microfluidics: Experimental Optimization of Droplet Size and Generation Time through Flow Focusing, Phase, and Geometry Variation

Adam Britel*[a], Giulia Tomagra[b], Pietro Aprà[a], Veronica Varzi[a], Sofia Sturari[a], Nour hanne Amine, Paolo Olivero[a], Federico Picollo[a]

Droplet-based microfluidics systems have become widely used in recent years thanks to their advantages, varying from the possibility of handling small fluid volumes to directly synthesizing and encapsulating various living forms for biological-related applications. The effectiveness of such systems mainly depends on the ability to control some of these system's parameters, such as produced droplet size and formation time, which represents a challenging task. This work reports an experimental study on tuning droplet size and generation time in a flow-focusing geometry fabricated with stereolithography 3D printing by exploring the interplay of phase and geometrical parameters. We produced droplets at different low flow rates of continuous and dispersed phases to assess the effect of each of these phases on the droplets' size and formation time. We observed that smaller droplets were produced for high viscosity oil and water phase, along with high flow rates. In addition, changing the microfluidics channels' width, and morphology of the orifice has shown a similar effect on droplet size, as shown in the case of high-viscosity solutions. The variation of the bifurcation angle shows a noticeable variation in terms of the achieved droplet size and formation time. We further investigated the impact of modifying the width ratio of the continuous and dispersed phases on droplet formation.

## 1 Introduction

Droplet-based microfluidic systems have risen as important tools in recent years, offering various advantages ranging from the encapsulation of single living forms for biological-related applications[1] to chemical syntheses[2,3] and many other applications[4,5]. Droplet microfluidics allow the manipulation of small volumes of fluids in immiscible phases, usually water and oil, with low Reynolds numbers. In these systems, mainly two key components can be identified, namely: the dispersed phase which refers to the solution to be encapsulated, and the continuous phase, which refers to the medium on which the droplets flow. The primary benefit of using droplet microfluidics is represented by the ability to control the encapsulation of small volumes of fluids, in the order of femto- to nanolitres. In this context, the capability of controlling the generated droplet size emerges as a key requirement. Droplets in droplet-based microfluidics can be generated either through active methods or passive methods. Active methods involve the use of external sources, such as electric, optical, and thermal control to rupture the liquid-liquid interface into droplets[6–8]. In contrast, passive methods take advantage from the surface properties of both the continuous and dispersed phases along with the channel geometry to produce droplets[9]. When it comes to passive droplet microfluidics, there are several designs available, but the most frequently employed ones are the use of T-junctions and flow focusing, each with its specific droplet formation processes[10,11]. Flow-focusing geometries stand out from other geometries due to their adaptability in terms of creating droplets of varied sizes, generation rates, and a lower coefficient of variation in droplet size[9–13]. However, the efficiency of these systems depends mainly on the ability to control key properties such as droplet size and formation time, which as previous studies have shown can be tuned by controlling device parameters[14–18]. For example, in biological applications, the precise control over droplet size is an integral aspect of optimizing encapsulation efficiency and consistency[19]. Small droplets are preferred for single-cell-related analysis while larger ones offer ease of manipulation which is essential for droplet-sorting-related applications and multiple reagents mixing[4]. Correspondingly, the droplet formation time directly affects the overall efficiency of droplet production for the desired use.

The methods used for device fabrication impose constraints on the geometric features of the microfluidic channels, which, as discussed above, determine the performance of droplet generation. Traditional fabrication methods to develop microfluidic devices, namely soft lithography, are often labor-intensive and time-consuming. With the

[a.] Department of Physics and "NIS" Inter-departmental Centre, University of Torino National Institute of Nuclear Physics, sect. Torino, Via Pietro Giuria 1, 10125 Torino, Italy.
E-mail: a.britel@unito.it.

[b.] Department of Drug and Science Technology, NIS Interdepartmental Centre, University of Torino, Corso Raffaello 29, 10125 Torino, Italy.

use of Polydimethylsiloxane (PDMS) in soft lithography, good control over channel dimensions is achieved with cost-effective approach, but the fabrication method is two-dimensional, and thus the production of three-dimensional structures requires multiple steps. For such a reason, 3D printing emerged as a solution for accelerating microfluidics prototypes development and devices production. In the present work, it allowed the introduction of various features in the same device allowing the design of customized geometries for specific applications. Moreover, the ease of use, low cost of fabrication, and the possibility of multiple applications of the same device make it an attractive option on the industrial and the research scale. The resolution of the produced devices depends on the used 3D printer and technique for the fabrication that usually associates high quality with the cost of equipment. In this work, we focus our study on the flow-focusing configuration, more specifically, we use parallel dispersed phase channels with a contraction in the continuous phase channel. In this study, we used traditional quasi 2D-flow focusing design to validate our findings, to assess the practicality of using desktop 3D printers, and to enhance the accessibility of our findings to the broader research community**.** We present an experimental investigation on the tuning of droplet size and formation time produced within the stereolithography (SLA) 3D-printed flow-focusing geometry, aiming to explain the interplay between the solution-related and geometrical parameters. In order to present results of interest to a wider audience within the microfluidics community, we limited our research to microfluidic channels of a few hundred microns in diameter, which can be achieved with standard desktop 3D printers. This work aims to shed light on the possibility of obtaining small droplets with a controllable formation time by adjusting different geometric and solution-related parameters. By understanding these factors, our work not only can open new possibilities to precisely controlling and optimizing droplet formation processes but also contribute to the fundamental knowledge base, enhancing the understanding and accessibility of this field for a broader scientific audience. To facilitate the observation of individual droplets and more easily understand the impact of the parameters under study, we worked at low flow rates, both for continuous and dispersed phases.

## 2 Materials and methods

### 2.1 Experimental setup

All the microfluidic devices developed in this work were manufactured with an SLA 3D printer from Formlabs (Form 3B) using Formlabs clear resin. We used only flow-focusing geometry as it allows to precisely adjust the size of the generated droplets by controlling the flow rates. The fabrication process starts by designing the device using any computer-aided design (CAD) program: in this work, we use Autodesk Fusion 360. All devices in this study were printed at a 50 μm resolution directly on the printer build platform. The devices were printed with clear resin because its transparency made it simpler to monitor the droplet formation inside the microfluidic channels. Once the device was ready, it was cleaned with Isopropyl Alcohol (IPA) of a concentration equal or higher than 99 % to remove the uncured resin on the surface and inner channels of the microfluidic devices. The printed devices were agitated for 5 minutes in IPA, followed using pressurized air to force the removal of stuck resin from the channels. Then, IPA was pumped inside the channels to ensure the absence of any remaining stuck resin. The last step is the most important as poor cleaning can lead to blocked channels. Once the cleaning was performed, the device was post- cured in a UV chamber to ensure complete polymerization and to optimize material properties to their full potential. The size of the microfluidic channels has been kept in the order of 500 μm to 600 μm in order to keep our fabrication results achievable with a standard desktop 3D printer. Each printed device was rinsed with the oil that will be employed a few hours before use, which helped to reduce clogging caused by droplets sticking to the walls of the microfluidic channels. This operation allowed for the rinsed device to be used continuously for 4 to 5 hours without showing signs of cohesion, for the fluids employed in this work. In the section of this work dedicated to the influence of the solution-related parameters, we utilized distilled water and cell cultural medium (RPMI-1640) for the dispersed phase. The composition of RPMI-1640 medium is the following 10% horse serum (Invitrogen), 5% fetal bovine serum (Invitrogen) and 2% antibiotic/ antimitotic (pen/strep Invitrogen), that is usually used for the culture of PC 12 cells line at a temperature of 37 °C in a 5% CO2 atmosphere. For the continuous phase, we used paraffin oil, heavy mineral oil, castor oil, and a combination of heavy mineral oil with a surfactant (Span 80, Sigma-Aldrich). In the section of this work related to the influence of the geometrical parameters, we kept the distilled water as a dispersed phase and heavy mineral oil as a continuous phase. Figure 1 shows flow diagrams of the various experimental conditions considered for the evaluation of the impact of the solution-related and geometric parameters.

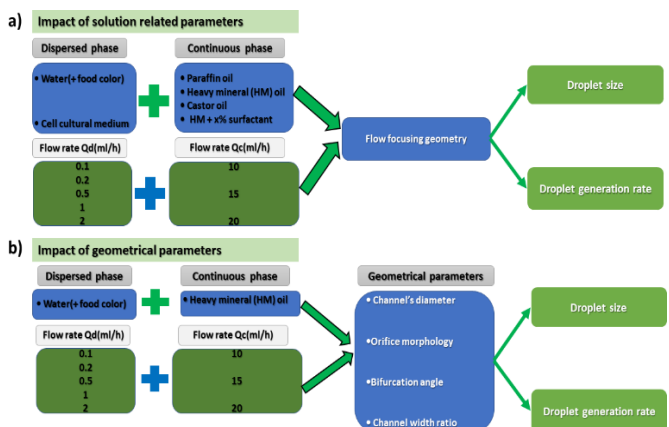

**Fig. 1** Flow charts representing the various parameters examined to assess their influence on the droplets size and formation time. 1.a) case of the solution related parameters and 1.b) case of the geometrical related parameters.

Dual syringe pump (Harvard Apparatus Model 33 Twin Syringe Pump) was employed to infuse/withdraw two liquids in the microfluidic devices. The flow rates of the dispersed phase ($Q_d$) varied between 0.1 ml/h to 2 ml/h while for the continuous phase ($Q_c$) they were set between 10 ml/h and 20 ml/h. Values of $Q_d$ and $Q_c$ limited to these ranges allow the formation of single droplets, which is the main interest of the present work. The experimental setup is composed of a syringe pump, a light source, and a camera capable of capturing 30 fps (SVPRO 4K), as shown in Figure 2. The tubes were directly inserted into the microfluidic devices without the need of additional connectors which helps reducing the risk of leakage. Images of the produced droplets were extracted from the recorded video and analyzed by the open-source image processing software ImageJ[20]. Due to the slow generation rate of droplets, we measured three distinct droplets at different timing during the experiment. This approach was used to assess the repeatability and precision of single droplet measurements. In all of the presented graphs, the error bars are not visible because their size coincides with the smallest possible size of the graph points. These error bars represent the variability observed in these repeated measurements of droplets rather than polydispersity over a large population, providing insights into the consistency of our droplet formation process. However, statistical analysis was conducted to ensure the reliability of the presented data. All the experiments were performed in the same environmental conditions to avoid any undesired effect that may arise from the change of temperature, for example.

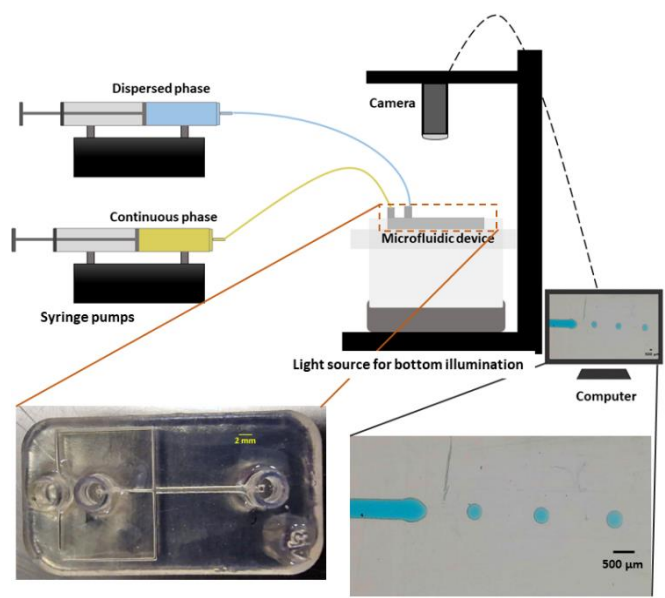

**Fig.2** Illustration of experimental setup utilized for the record and the analysis of the formed droplets. Bottom left depicts a 3D printed microfluidic device used for droplets generation, and bottom right shows an image of the generated droplets.

## 3 Results and discussions

### 3.1 Investigation of solution related parameters on droplet size and formation time

**Effect of flow rate**

In this section, we used rectangular channels of 600 µm width and 1 mm height. To assess the effect of the flow rate on the droplet size and formation time, we utilized distilled water as the dispersed phase and heavy mineral oil as the continuous phase. The experiment consists of varying the flow rate of the dispersed phase between 0.1 ml/h and 2 ml/h for each of the three selected flow rates of the continuous phase, i.e. 10,ml/h and 20 ml/h. Figure 3.a illustrates the device that was used for these tests, and 3.b and 3.c shows the results graphs for the assessment of the influence of the flow rates on the droplet size and formation time respectively. By increasing the flow rate of the dispersed phase ($Q_d$), an increase in water-in-oil droplet size was observed for all continuous phase flow rates ($Q_c$) used. These results show that the increase of $Q_d$ led to an increase in the fluid velocity across the microfluidic channel resulting in a higher sheer stress at the oil-water interface. This increases the pressure drop, which acts as a driving force and reduces the surface tension at the interface between the two immiscible phases, resulting in the formation of larger droplets. Droplet formation time is also affected by the change in $Q_d$, whose increases favoring faster droplet formation. In fact, it is assumed that increasing the flow rate of the dispersed phase increases the shear stress at the interface between the two phases, which helps to overcome the cohesive forces within the dispersed phase, resulting in faster droplet formation. Khorrami et al[22] explained these results in terms of capillary and Weber numbers. Concerning the effect of $Q_c$, when we increased this parameter from 10 ml/h to 20 ml/h we observed that for the same $Q_d$ value(s), from 0.1 ml/h to 2 ml/h, the size of the droplet drastically decreased. The continuous phase is surrounding the dispersed phase, thus as the $Q_c$ increases, the continuous phase tends to occupy more space in the main channel leaving less space for the dispersed phase to go through it. This change promotes the breakup of the dispersed phase into smaller droplets. Therefore, the increase in $Q_c$ has led to an increase in the shearing force resulting in smaller droplets[23]. For what concerns the formation time, a similar effect to the case of the influence of the flow rate of the dispersed phase was observed.

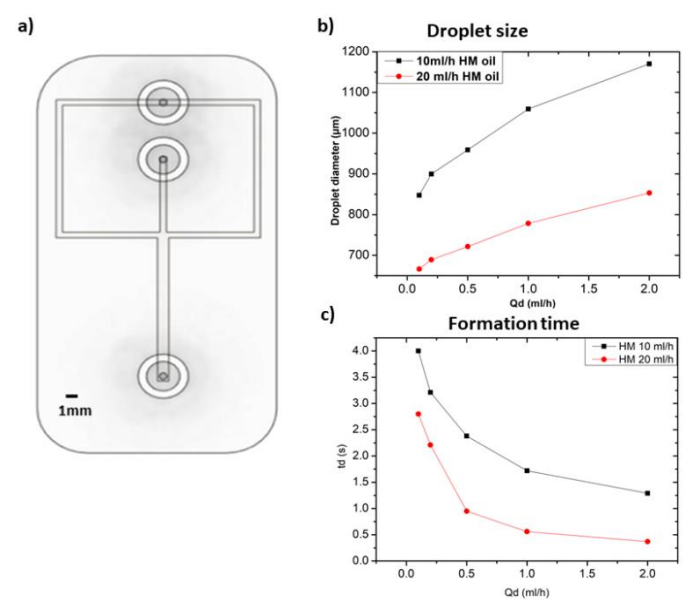

**Fig.3** a. CAD illustration of the flow focusing microfluidic device utilized for the solution related parameters experiments. Both channels of the continuous and dispersed phase have a width of 600 µm while the collecting channel is 1 mm. b. Influence of the flow rate of the continuous and dispersed phase on the droplet size. c. Influence of dispersed and continuous phase flow rates on droplets formation time.

**Effects of continuous phase viscosity**

The interaction between the employed fluids, their viscosities and their mutual interfacial tension are the primary determinants controlling droplet formation in the microfluidic channel[23]. Thus, this subsection aims to highlight the role of the viscosity of the continuous phase on the size and formation time of droplets. For the following tests, we employed the distilled water as the dispersed phase and three oils as the continuous phase, namely: paraffin oil, heavy mineral oil, and Castor oil with viscosities of 25-80 mPas, 80 mPas, and 600-800 mPas at 20 °C respectively. Figure 4 illustrates the effects on droplet size and formation time upon changing the viscosity of the employed continuous phase. Figure 4.a shows that for an increase in the viscosity of the continuous phase, there was a considerable decrease in the size of the droplets. It should also be noted that increasing the viscosity of the continuous phase, e.g., castor oil, together with a high $Q_c$ and low $Q_d$, can result in tiny droplets compared to a low viscosity fluid, i.e., Paraffin oil as illustrated in Figure 4.a. The same results were observed in the case of T-junction geometry, i.e., a microfluidic geometry where the two fluid channels intersect at a 90-degree angle, resembling the shape of letter "T", meaning that these results are independent of the adopted geometry[24]. In addition, the change in the viscosity ratio between the continuous and dispersed phases leads to other modifications related to the shape of the formed droplets, which are beyond the scope of this paper but have been discussed by Srikanth et al[23]. Moreover, the decrease in droplets diameter vs. $Q_d$ variation was approximately linear in the case of low viscosity oil, while it was much less for higher viscosity oil. The droplet formation time is significantly influenced by the change in the viscosities. At higher viscosities of the continuous phase, the formation of droplets got faster, as illustrated in Figure 4. b. Increasing the viscosity of the continuous phase led to an increase in the viscous shear stress acting at the interface between the two immiscible fluids, resulting in a significant acceleration of droplet formation[25]. At higher viscosities, the variation in formation time was minimal as $Q_c/Q_d$ ratio decreased. Hence, increasing both of the viscosity of the continuous phase and the viscosity ratio between the two phases resulted in a more significant impact on the formation time of droplets with respect to varying the flow rates.

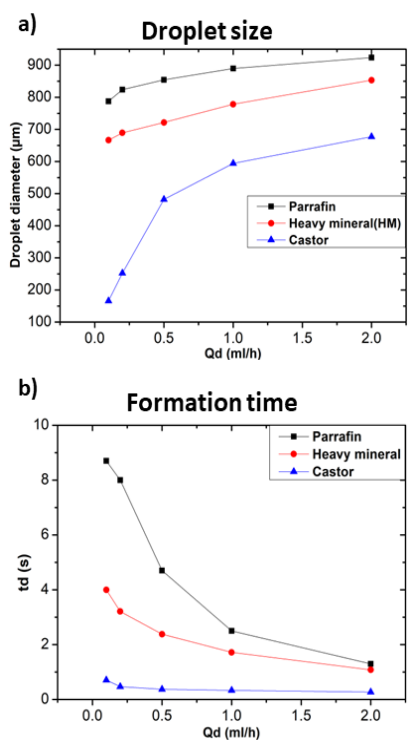

**Fig.4** Influences of continuous phase viscosities at $Q_c$ = 20ml/h. a) effect on droplet size at different $Q_d$. b) effect on droplet formation time at different $Q_d$

### Interfacial tension effects

The fluids used as continuous phases influenced the size and formation time of the produced droplets, since they have a specific viscosity, as seen in the previous section, but they can also have a different interaction with the dispersed phase. In order to modify the interfacial tension between the two fluids, heavy mineral oil with added specific concentrations of surfactant (Span 80, Sigma-Aldrich), ranging from 1.0% to 2.5% of the total weight (wt), was used. In the present work, wt was limited to 2.5% because of the clogging and droplet formation problems observed above this level. In this part, distilled water was used as the dispersed phase. Surfactants are substances that form self-assembled molecular clusters, known as micelles, in a solution. They also act at the interface between two phases, reducing the interfacial tension and facilitating the mixing or separation of the phases[26]. In the described experimental configuration, the surfactant primarily acted at the interface between the continuous and dispersed phases, by reducing the interfacial tension of the droplets and lining their inner surfaces with a hydrophilic layer. Surfactants are often used to stabilize droplets and prevent the merging of droplets that happen to enter in close contact[27,28]. Figures 5.a and 5.b shows the effects of the concentration of the surfactant on the size and formation time of droplets.

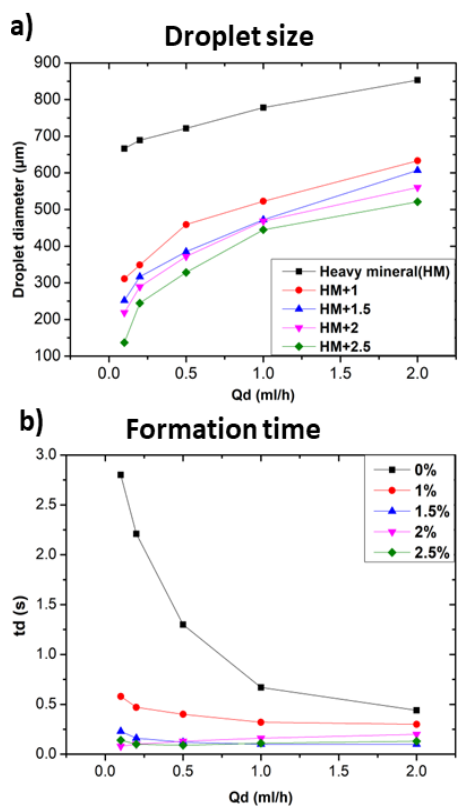

**Fig.5** Influence of the composition of the dispersed and continuous phase on the droplets size and formation at $Q_c$ = 20ml/h. 5.a) and b) shows the influence of the continuous phase composition in terms of a) droplet size, b) formation time.

In particular, Figure 5.a. shows that the addition of 1% wt Span 80 to heavy mineral oil drastically reduced the droplet size with respect to the case of only heavy mineral oil, while increasing the surfactant concentration resulted in a relatively smaller decrease of the droplet size. These experiments proved the impact of the interfacial tension on the droplet's formation process, i.e. at lower interfacial tension smaller droplets were observed[29]. All curves characterised by a different surfactant concentration versus the variation of the dispersed flow approximately follow the same trend. At 1 % wt, the surfactant molecules reduce the interfacial tensions effects resulting in an easier the breakup of water-in-droplets which significantly reduces the sizes of the droplets. At higher concentrations, i.e., 2.5% wt, the size of the droplets keeps slowly decreasing since is approaching the limit at which the interfacial tension can be lowered. The evolution of the droplet formation time at different surfactant concentration follows the same trend as the one of droplets size as described above. However, these results cannot be generalized to all kinds of surfactants, as this effect depends on their compatibility with the used phase[30]. A comparison of the droplet size obtained using continuous phases of different viscosities with respect to heavy mineral oil, including various concentrations of surfactant, is shown in Figure 6. A flow rate ratio of 100 was selected, corresponding to $Q_d$ = 0.2 ml/h and $Q_c$ = 20 ml/h. Notably, using castor oil (i.e. high viscosity regime) as the continuous phase or supplementing heavy mineral oil with 2.5 wt% surfactants yielded to comparably sized droplets. This confirms that the reduction of droplet size can be achieved through two distinct mechanisms, namely either by modulating the interfacial tension via surfactant addition or by increasing the viscosity of the continuous phase.

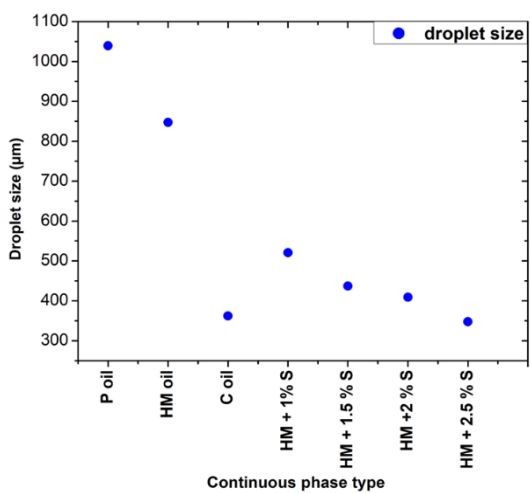

In the following the effect of changing the composition of the dispersed phase while keeping heavy mineral oil as the continuous phase was investigated. Given the general interest in the development of droplet microfluidic systems for cell encapsulation, a standard medium for cell culture (RPMI-1640, Invitrogen) was selected as dispersed phase and compared with distilled water.

**Fig.6** Influence of continuous phase viscosity and concentration of surfactant on the droplet size at a flow ratio of 100. P refers to Paraffin, HM to heavy mineral, S to surfactant, C to Castor

Figures 7.a and 7.b show that cellular medium improved the performance of the microfluidic systems resulting in the production of smaller droplets and faster formation time in comparison to the case of distilled water. Different factors such as changes in viscosity, surface tension, or interfacial properties affect the droplet formation. Distilled water and cell medium are likely to have distinct surface tensions with the used continuous phase resulting in a modification of the droplet formation mechanism. As we experienced a decrease in the droplet size, we assumed that the modification of the dispersed phase caused a decrease in the surface tension[31]. In terms of interfacial properties, it is most likely that the composition of the used cell medium is more compatible with heavy mineral oil, contrarily to what obtained with distilled water. Hence, the observed change concerning the size and formation time of the produced droplets.

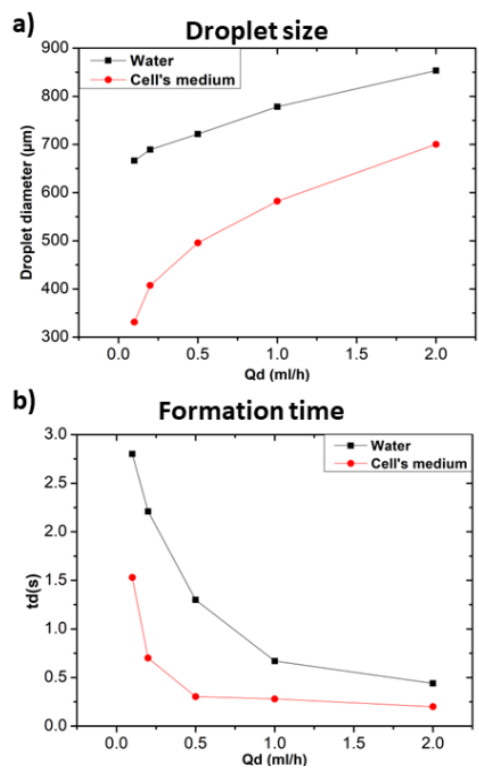

**Fig.7** Influence of the dispersed phase composition at $Q_c$ = 20ml/h in terms of a) droplet size, b) formation time

### 3.2 Investigation of geometrical parameters on droplet size and formation time

**Channel width influence**

In addition to the previously discussed solution-related parameters, this section aims at highlighting the influence of various geometrical parameters in achieving controllable droplet size and formation time. The existence of different geometries for the generation of droplets itself is proof that the geometry of the device plays a primary role in determining the produced droplets' geometrical characteristics[32]. Channel width can have a significant influence on the formed droplet properties as it directly affects the fluid flow dynamics within the device[11]. In Figure 8, experimental results from three different microfluidics channel widths (i.e. 500 µm, 600 µm, and 700 µm) and their effects on droplets size and formation time are presented. The microfluidic devices used in this subsection are designed with a rectangular cross section, with a constant height of 1mm and varying width. Distilled water and heavy mineral oil were used respectively as the dispersed continuous phases for all experiments discussed in this subsection. As expected, larger channel widths lead to the formation of larger droplets compared to those formed in channels with smaller widths. This is because the larger cross-sectional area allows for reduced shear stress at the fluid interface, causing droplet break-off to occur more slowly and thus leading to larger droplets[9]. Additionally, the larger cross-sectional area also impacts the fluid pressure, which in turn influences droplet size. Specifically, changes in cross-sectional area further affect the rate at which droplets break off and their resulting size[32]. In terms of formation time, droplets require more time to be formed in the case 700 µm width channel with respect to the ones in 500 µm channels, as shown in Figure 8.b. In larger channels, a longer distance must be travelled by the fluid to reach the droplet break-off point. This path increase, along with the decrease in shear stress, resulted in longer droplet formation times. Similar results were observed in the case of T-junction geometry[33]. Nevertheless, the influence of the modification of the channel width cannot be limited to these effects, since the increase of the channel width allows also for larger flow rates, thus affecting the interaction between the continuous and dispersed phases and resulting in a modified droplet formation process.

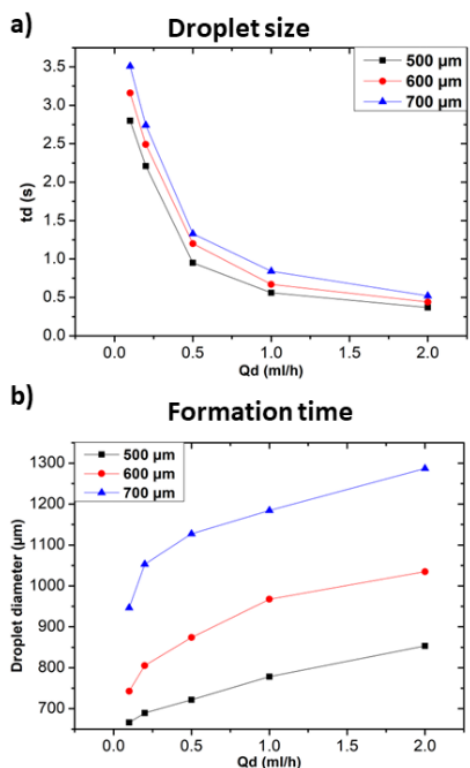

**Fig.8** Influences of channel's width at $Q_c$ = 20ml/h. a) effect on droplet size at different $Q_d$. b) effect on droplet formation time at different $Q_d$.

**Influence of bifurcation angle**

Droplet microfluidic devices are characterised by the presence of a region, usually referred as a "junction", where the channel through which the dispersed phase flows meets the channels dedicated to the flow of the continuous phase. In this geometry, a bifurcation angle can be identified as the angle between the dispersed and continuous phase channels. To investigate the effect of the bifurcation angle on droplet formation, three microfluidic devices with the same channel diameter and three different bifurcation angles (i.e. 45°, 90° and 135°) were fabricated as shown in Figure 9.a.

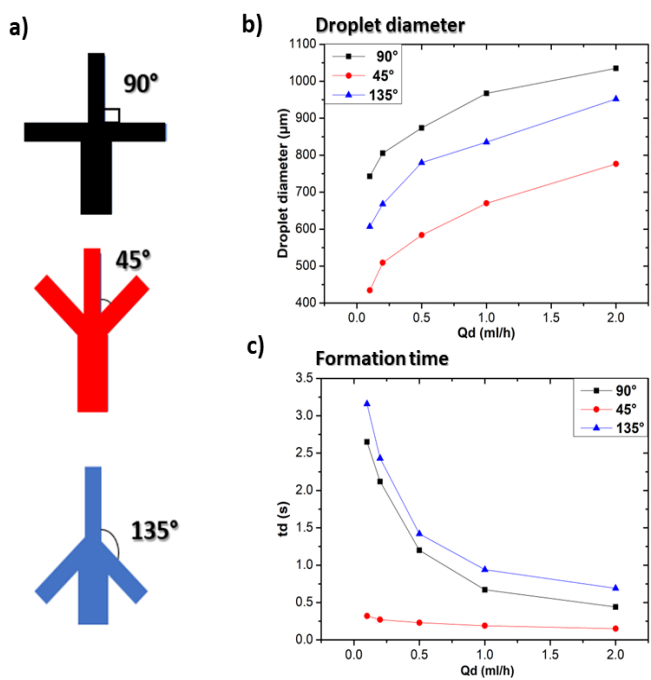

**Fig.9** Influence of bifurcation angle at $Q_c$= 20ml/h. a) Graphical illustration of the three used bifurcation angles, b) effect on droplet size at different $Q_d$, c) effect on droplet size at different $Q_d$.

Figure 9.b shows a significant influence of the bifurcation angle on the formed droplet size: for example, at $Q_d$ = 0.1 ml/h droplets formed using the 45° configuration were 28% times and 52% times smaller than those obtained using an angle of 135°and 90° respectively. This difference can be explained by a change in the flow dynamics since the increase of the bifurcation angle yields an increase in the shear stress applied to the dispersed phase, resulting in larger droplets[11]. In the case of a 90° angle, the flow paths of the continuous and dispersed phases meet perpendicularly to each other. This perpendicular intersection amplifies the shearing forces acting on the fluids, leading to the formation of larger droplets, while for 45° a decreased shear stress is expected. In terms of the formation time, the variation of the bifurcation angle significantly influences the droplet formation time, thus providing an additional level of control in microfluidic droplet generation systems. At a 45° angle, we observed the quickest droplet generation process, as illustrated in Figure 9.c. In this configuration, the dispersed phase encounters a more abrupt change in direction, which facilitates faster droplet pinch-off. Also, the variation of the formation time with $Q_d$ is less notable as in the case of the other two angles. The results of this subsection outline that the flow dynamics in the bifurcation angle play an important role in determining droplet size and formation time. Nevertheless, it is worth remarking the role of the bifurcation angle can also alter the droplet formation regime[34] and the peak internal fluid velocity within individual droplets[35].

**Influence of orifice morphology**

The small opening that connects the continuous and dispersed phase channels to the collecting channel (usually referred as the "orifice") has an impact on droplet size and formation time. Devices with five different orifices geometries were 3D-printed to examine the impact of orifice geometry on droplet size and formation time. Adopted geometries and collected data used for this subsection are illustrated in Figure 10.a. Figure 10.b shows the effect of the orifice morphology on the formed droplet size. In the case of the square orifices, namely 500×500 µm², 400×400 µm² and 300×300 µm², the droplet size decreases with the size of the section of the orifice. These results are consistent with the trend of droplet formation in microfluidic flow focusing geometries[11,36,37], since a small orifice provides less space for the dispersed phase to assemble and pinch off. This effect can alter droplet dynamics, which can also be influenced by factors such as the capillary number[23]. For the case of rectangular orifices, (namely 500 × 200 µm² and 500 × 100 µm²), the observed droplets are smaller than the ones relevant to the square orifices. This result might appear as counterintuitive at first glance. Certainly, SLA 3D printing allows precise control over the geometry and dimensions of the microfluidic devices produced. However, smaller orifices can pose some problems in terms of post-processing, particularly in terms of cleaning and removing the uncured resin inside the channels. The narrow confines of the small orifice make the removal of the uncured resin inside the channels challenging and most often results in blocked channels, resulting in non-functional microfluidic devices. This is particularly true for low-cost desktop 3D printers, while the more expensive ones can more easily achieve even smaller features. Orifice with asymmetrical aspect ratio (while keeping the section area constant) are easier to clean with respect to symmetrical ones. For example, the success rate on the fabrication process using 500 × 200 µm² orifice was higher with respect to 300 × 300 µm² orifice (90% vs 20%, respectively). The 500 µm width makes the channel more accessible, thus allowing for a more efficient cleaning, and a simplified manufacturing process. The purpose is to highlight that small geometrical modifications can help improving the devices reproducibility and avoiding printing issues caused by resin stuck in the channels. In terms of formation time, for the case of square orifices, the droplet formation time decreases as the orifice size increases. In general, the orifice provides hydrodynamic resistance to the flow of the dispersed phase. For the dispersed phase to pass through the orifice and form a droplet, it has to overcome this resistance. This hydrodynamic resistance is intrinsically linked to the geometrical properties of the orifice, especially its size[38]. Consequently, the droplet formation time inversely scales with the orifice section area, as demonstrated in Figure 10.c. Nevertheless, the shape and dimensions of the orifice, often referred to as its aspect ratio, play a crucial role in determining how quickly droplets form and separate from the main flow[39].

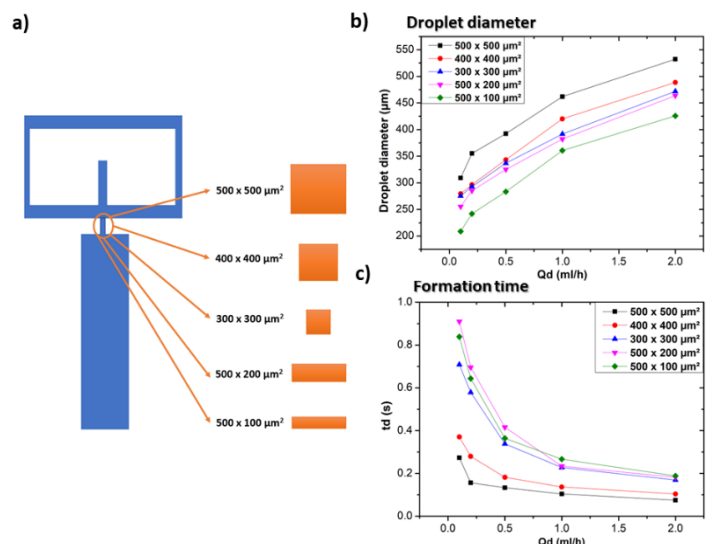

**Fig.10** Influence of orifice morphology at $Q_c$ = 20 ml/h. a) Graphical illustration of the various forms of orifices employed in this subsection, b) impact on droplet size at different $Q_c$, c) impact on droplet formation time at different $Q_c$

**Influence of channel width ratio**

To study the impact on droplet size and formation time based on different channel dimensions for continuous and dispersed phases, two devices were created. The first device, labelled as "Water 800", has a dispersed phase channel width that is double with respect to the continuous phase channel. Conversely, the second device, labelled as "Oil 800", features a continuous phase channel width that is double in size with respect to the width of the dispersed phase channel. Both devices are illustrated in Figure 11.a. Figure 11.b shows the effect of changing the channel width ratio on droplet size.

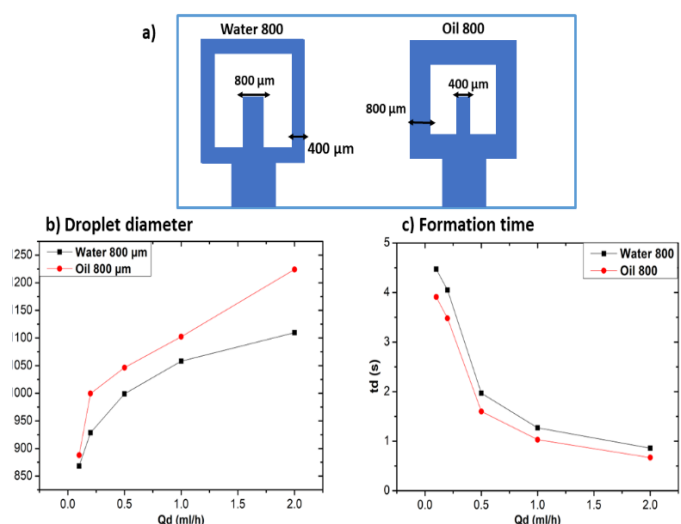

**Fig.11** Influence of channel width ratio at $Q_c$ = 20 ml/h. a) graphical illustration of the geometry of the two devices of this subsection, c) influence on droplet size at different $Q_d$, c) influence on droplet formation time at different $Q_d$.

The Oil 800 units produced larger droplets compared to Water 800. In the case of Oil 800, the large width of the continuous phase channel allows a greater volume of continuous phase to interact with the dispersed phase at the droplet formation site. As a result, a greater volume of the dispersed phase can be encapsulated, resulting in larger droplets. Similar considerations can be made for the difference in formation times reported in Figure 11.c.

## 4 Optimized geometrical combination for minimized droplet production.

After examining how various geometric and solution related parameters impact droplet size, this section aims to shed light on the combined effect of these geometric parameters to achieve the smallest droplet size. We maintained the microfluidic channel width in the 500 – 600 µm range and a height of 1 mm to ensure that it can be reproduced by using any standard desktop 3D printer. Distilled water and heavy mineral oil were respectively used as the dispersed and continuous phases to easily compare these results with the ones shown in Figure 3. The design of this device that integrates all the geometric features discussed in the previous section (i.e. orifice morphology, bifurcation angle, and channel width) is depicted in Figure 12.a. The observed droplet size is presented in Figure 12.b. As anticipated, a noticeable reduction in droplet size is observed as compared to the previous results. In fact, for $Q_d$ = 0.1 ml/h and $Q_c$ = 20 ml/h, the droplet size becomes approximately 90% smaller with respect to the same parameters in Figure 3. This shows the effectiveness of systematic understanding of the effect of device geometry on droplet size, thus facilitating the production of smaller droplets in a quick, simple, and cost-effective way.

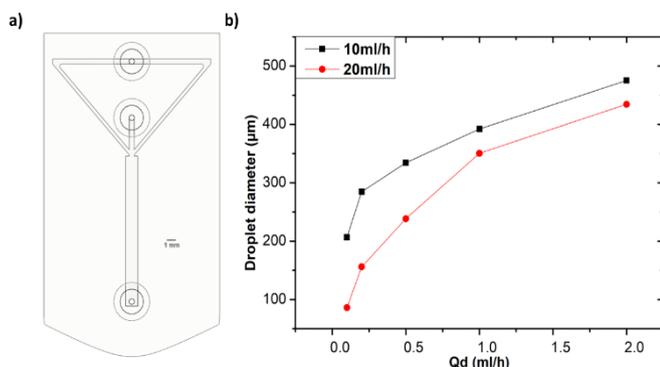

**Fig.12** a. CAD illustration of the optimized microfluidic device utilized for this subsection experiments. The width of the channels of continuous and dispersed phase were kept to 600 µm. b. recorded droplet size at different $Q_d$.

## 5 Conclusion

This work offers a systematic experimental investigation of the different parameters influencing droplet size and formation time in 3D printed microfluidic devices, reporting on the interplay between the flow rates of the used phases, viscosity and composition of the used fluids, the addition of surfactants, and geometric parameters of the microfluidic device. 3D printing was selected as a tool for rapid prototyping allowing the investigation of multiple geometric parameters in the droplet formation process. Through this work, we have demonstrated the effective use of desktop 3D printer in the field of droplet microfluidics, thus highlighting their capability to achieve an accurate control over droplet characteristics under accessible conditions. In our experiments we systematically investigated various parameters, leading to the generation of droplets as small as 80 µm. This represents a remarkable result, considering the simplicity and affordability of the used equipment. The key findings of our study reveal that careful manipulation of flow rates, fluid viscosities, and geometric parameters of the employed microfluidic device can finely tune droplet size and formation time. In particular, this work shows the role of channel width, and orifice morphology, providing a path towards optimizing droplet characteristics even with the constraints of affordable desktop 3D printers. Our results resonate with the findings of advanced research in this field which shows that complex microfluidic tasks, typically reliant on specific equipment, can be replicated to a certain extent using more accessible technology. Overall, these results provide practical strategies for manipulating droplet size and formation time within the constraints of affordable 3D printing methods. The comprehensive exploration of the diverse solution-related, and geometrical parameters offers valuable insights for the design and operation of droplet microfluidic systems within the constraints of limited facilities.

## Author Contributions

A.B conducted the formal analysis, investigation, methodology and drafted the manuscript. G.T provided resources for experiments and validated the procedures. P.A performed formal analysis. V.V., S.S., and N.A were involved in the validation and review of the first draft. P.O was responsible for manuscript review and editing. F.P conceptualized the study, provided supervision, created visualizations, and contributed to manuscript drafting. All authors have read and agreed on the content of the manuscript.

## Conflicts of interest

The authors declare that they have no conflicts that may influence the work reported here.

## Acknowledgements

This research has received funding from the European Union's H2020 Marie Curie ITN project LasIonDef (GA n.956387), and "QuantDia " project funded by the Italian Ministry for Instruction, University and Research within the " FISR 2019 ''program.